\documentclass[preprint,showpacs,amsfonts,aps,superscriptaddress,floatfix]{revtex4}
\usepackage{epsfig}
\usepackage{bm}

\chardef\atcode=\catcode`\@
\catcode`\@=11
\@addtoreset{equation}{section}
\renewcommand{\theequation}{\arabic{section}.\arabic{equation}}

\def\Re{{\rm Re}}
\def\Im{{\rm Im}}
\def\be{\begin{equation}}
\def\ee{\end{equation}}
\def\bea{\begin{eqnarray}}
\def\eea{\end{eqnarray}}

\begin{document}

\title{Impact of the QCD four-quark condensate on in-medium spectral
changes of light vector mesons}

\author{S. Zschocke}
\affiliation{Forschungszentrum Rossendorf, PF 510119, 01314 Dresden, Germany}
\author{O.P. Pavlenko}
\affiliation{Institute for Theoretical Physics, 03143 Kiev - 143, Ukraine}
\author{B. K\"ampfer}
\affiliation{Forschungszentrum Rossendorf, PF 510119, 01314 Dresden, Germany}

\begin{abstract}
Within the Borel QCD sum rule approach at finite baryon density we study 
the role of the four-quark condensates
for the modifications of the vector mesons $\rho$, $\omega$ and $\phi$
in nuclear matter.
We find that in-medium modifications of the $\rho$ and $\omega$ mesons are
essentially dominated by the dependence of the 4-quark condensate
on the nucleon density. In particular, the numerical value of a parameter ($\kappa_N$),
which describes the strength of the density dependence of the 4-quark 
condensate beyond the mean-field approximation, governs the decrease of the
$\rho$ mass as a function of the density. For the $\omega$ meson the sign of 
the in-medium mass shift is changed by variations of $\kappa_N$. To study
consistently the in-medium broadening of the light vector mesons we employ
$\rho N$ and $\omega N$ scattering amplitudes derived recently from a covariant
unitary coupled channel approach adjusted to pion- and photo-induced reactions.
In contrast to the $\rho$ and $\omega$ mesons, the in-medium mass of the $\phi$
meson is directly related to the chiral (strange) quark condensate.
Measurements of the vector meson spectral change in heavy-ion collisions
with HADES can shed light on the yet unknown density dependence of the
4-quark condensate. 
\end{abstract}

\maketitle

\section{Introduction}

Changes of the vector meson properties in strongly interacting matter 
at finite baryon density and temperature are presently of great interest,
both theoretically and experimentally. In particular, the current 
heavy-ion experiments with the detector HADES \cite{lit1} at the
heavy-ion synchrotron SIS18 
(GSI, Darmstadt) are mainly aimed at measuring in-medium modifications 
of light vector meson via the $e^{+} e^{-}$ decay channel with high 
accuracy. One of the primary goals of the future experiments planned at 
SIS100/200 is also to study very dense baryon matter and the
expected strong changes of the in-medium hadrons. 

It is widely believed that the in-medium spectral change of the light mesons is 
related to the chiral symmetry restoration at finite temperature and 
baryon density. There are indeed various theoretical indications 
concerning an important sensitivity of the meson spectral density on the 
partial restoration of the chiral symmetry in a hot/dense nuclear medium. 
For instance, at finite temperature the vector and axial-vector meson 
correlators become mixed in accordance with in-medium Weinberg sum rules    
\cite{lit2,lit3}. 
Such a mixing causes an increasing degeneracy of
vector and axial-vector spectral functions which would manifest themselves 
as a decrease 
of the $\rho$ and $a_1$ meson mass splitting. 
Similarly, the degeneracy of  
scalar ($\sigma$ channel) and 
pseudo-scalar ($\pi$ channel) correlators found 
in lattice QCD \cite{lit4} can lead to a considerable enhancement of the 
$\sigma$ meson spectral function at finite temperature and density 
\cite{lit5}.

In spite of substantial efforts undertaken to understand the nature 
of vector mesons in a dense medium there is so far no unique and widely 
accepted quantitative picture of their in-medium behavior. 
The Brown and Rho conjecture \cite{lit6} on the direct interlocking  
of vector meson masses and chiral quark condensate $\langle\overline{q} q
\rangle_n$ 
supplemented by the "vector manifestation" of chiral symmetry in medium 
\cite{lit7,lit8} predict a strong and quantitatively the same 
decrease of the in-medium $\rho$ and $\omega$ meson masses. 

At the same time, model calculations based on various effective Lagrangians 
(cf. \cite{lit9}) predict rather moderate and different 
mass shifts for $\rho$ and $\omega$ mesons in a dense medium. In order 
"to match" both sets of predictions one has to go beyond simplifications made 
in the above mentioned approaches: The  
in-medium vector meson modification is governed not  
only by $\langle\overline{q} q\rangle_n$ but also 
by condensates of higher 
order to be evaluated beyond mean-field approximation. 
Further, effective Lagrangians  
are dealing with the scattering amplitudes in free space, but  
effects related to the in-medium change of the QCD condensates  
should be included \cite{lit10}.

The very consistent way to incorporate in-medium QCD condensates is 
through QCD sum rules (QSR). The QSR for vector mesons 
in nuclear matter were first developed in \cite{lit11}, 
where within a simple parameterization of the spectral density in terms 
of a delta function at the resonance peak an agreement with the Brown-Rho 
scaling, i.e. the same dropping of the $\rho$ and $\omega$ meson 
masses, in nuclear matter was obtained. While the zero-width approximation
for the resonance spectral density is successful in vacuum 
\cite{lit12}, such an approximation is not well grounded for the in-medium mesons which 
can undergo rather strong inelastic scatterings off the surrounding nucleons. For 
realistic in-medium QSR evaluations one needs to take into account the finite 
meson widths including collision broadening effects. 
The important impact of the 
finite width was studied, e.g., in \cite{lit13} using a plausible ansatz for the 
in-medium spectral density. As shown in this QSR analysis, there is no 
inevitable necessity for in-medium dropping of the vector meson masses, but 
the global changes of mesons like mass shift and width broadening turn out 
to be correlated in nuclear matter. To avoid too many unknown parameters in 
the QSR equation and to make more definite predictions one has to specify 
in a detailed manner the ansatz for the hadron spectral density. 
As we show below such a specification for $\rho$ and $\omega$ vector mesons 
can be done basing on an effective Lagrangian approach which gives a realistic 
behavior of the $\rho N$ and $\omega N$ scattering amplitudes.

As well known, QSR in nuclear matter contain also an uncertainty related to 
the poorly known density dependence of the four-quark condensate. The majority 
of the QSR evaluations employs mean-field approximations for the in-medium 
4-quark condensate, i.e. its density dependence is simply governed by the 
chiral condensate squared. At the same time, as pointed out in \cite{our1} 
the in-medium mass shift of the $\rho$ and $\omega$ mesons is dominated by the 
dependence of the 4-quark condensate on density. In particular, the sign of 
the $\omega$ meson mass shift is changed by the variation of the strength of 
the density dependence of the 4-quark condensate beyond mean-field 
approximation. This result was confirmed in \cite{our2}, where the $\omega$ 
meson spectral density was constrained within a general form of the in-medium 
$\omega$ meson propagator including collision broadening via the imaginary 
part of the $\omega N$ scattering amplitude delivered by an effective chiral 
Lagrangian \cite{lit14}.

A direct observation of the $\omega$ meson spectral change 
via the e$^+ e^-$ decay channel appears to be an experimental challenge 
in heavy-ion collisions at SIS18 energies. Both transport code simulations 
\cite{lit15} and a hydrodynamical model approach \cite{lit16} 
point to a considerable contribution of the reaction 
$\pi^+ \pi^- \rightarrow \rho \rightarrow e^+ e^-$
into dilepton spectra in the wanted region.
A chance to separate e$^+ e^-$ pairs from in-medium $\rho$ 
and $\omega$ mesons crucially depends on the quantitative details of their 
mass shift and width broadening in nuclear matter. This gives rise to a strong 
request from the experimental side to find out the $\rho$ and $\omega$ meson 
in-medium spectral changes simultaneously on a unique basis including 
self-consistently effects of the QCD condensates and collision broadening 
in nuclear matter.

In the present paper we study systematically within the Borel QSR the important 
role of the 4-quark condensate for spectral modifications of the $\rho$ 
and $\omega$ mesons in baryon matter. Being still within the low-density 
expansion we go beyond the mean-field approximation and vary the strength 
of the density dependence of the 4-quark condensate. Concerning the in-medium 
meson spectral density entering the hadronic part of the QSR evaluation we use 
a constraint motivated by the general structure of the vector meson 
propagator with finite in-medium width of $\rho$ and $\omega$ mesons 
reflecting the scattering of vector mesons off nucleons the in nuclear medium.
Seeking realistic $\rho N$ and $\omega N$ scattering amplitudes we 
employ the results of the recent covariant unitarized coupled
channel approach 
\cite{lit17} which satisfactorily 
describes the experimental pion- and photon-nucleon scattering data. 

We find that in-medium modifications of the $\rho$ and $\omega$ mesons are 
indeed dominated by the dependence of the 4-quark condensate on density. 
In particular, the numerical value of a parameter, which 
describes the strength of the linear density dependence of the 4-quark 
condensate, governs the decrease of the $\rho$ meson mass as a function of 
density. For the $\omega$ meson the sign of the in-medium mass shift is 
changed by variations of this parameter. Since the difference of 
the vector and axial-vector correlators is proportional to the 4-quark 
condensate the sign of the vector meson mass shift, measured via the 
$e^+ e^-$ channel, can serve as a tool for determining how 
fast nuclear matter approaches the chiral symmetry restoration with 
increasing baryon density.

Our paper is organized as follows. In section II we recapitulate the necessary
equations and formulate the Borel QCD sum rule. The systematic
evaluation of this sum rule is presented in section III for $\rho$ and $\omega$
mesons.  As supplement, we consider in section IV the case 
of the $\phi$ meson. The summary and a discussion can be found in section V.
Appendices A and B summarize the vacuum $\rho$ self-energy
and the $\rho, \omega$ meson-nucleon scattering amplitudes, respectively. 
In Appendix C we report on some technical details. 

\section{QCD sum rule equation}

For the sake of self-containment we list here the relevant equations
for the Borel QCD sum rule which our evaluations are based on.   

\subsection{Dispersion relation} 

Within QCD sum rules the in-medium vector mesons $V=\rho, \omega$ are 
considered as resonances  
in the current-current correlation function
\bea
\Pi_{\mu \nu} (q , n) = i \int d^4 x \;{\rm e}^{i q \cdot x} 
\langle {\cal T} \; J_\mu (x)\; J_\nu (0)\rangle_n\;,
\label{eq_5}
\eea
where $q_{\mu}=(E, {\bf q})$ is the meson four momentum, ${\cal T}$ denotes the 
time ordered product of the respective meson current operators 
$J_\mu (x)$,
and $\langle \cdots \rangle_n$ stands for the expectation value in medium. 
In what follows, we focus on the ground state of low-density baryon matter approximated 
by a Fermi gas with nucleon density $n$. 
We consider isospin symmetric nuclear matter, where the $\rho - \omega$ 
mixing effect is negligible \cite{lit18}. 
In terms of quark field operators, 
the vector meson currents are given by 
$J_\mu = \frac{1}{2} (\overline{\rm u} \gamma_{\mu} {\rm u} 
\mp \overline{\rm d} \gamma_{\mu} {\rm d})$, where the negative (positive) 
sign is for the $\rho$ ($\omega$) meson.
The correlator (\ref{eq_5}) can be reduced to  
$\frac{1}{3} \Pi_{\mu}^{\mu} (q^2, n) = 
\Pi^{(V)} (q^2, n)$ for a vector meson at rest, ${\bf q}=0$,  
in the rest frame of matter. 
In each of the vector meson channels the 
corresponding correlator 
$\Pi^{(V)}(q^2, n)$ satisfies the twice subtracted dispersion relation, 
which can be written with $Q^2 \equiv -q^2 = -E^2$ as 
\bea
\frac{\Pi^{(V)} (Q^2)}{Q^2} = \frac{\Pi^{(V)} (0,n)}{Q^2} - \Pi^{(V) '} 
(0) - Q^2 
\int\limits_0^{\infty} \; ds \frac{R^{(V)}(s)}{s (s + Q^2)}\;,
\label{eq_10}
\eea
with $\Pi^{(V)} (0,n) = \Pi^{(V)} (q^2=0, n)$ and $\Pi^{(V) '} (0)=
\frac{{\rm d} \Pi^{(V)} (q^2)}{{\rm d} q^2}|_{q^2=0}$ 
as subtraction constants, and 
$R^{(V)} (s) \equiv - \frac{1}{\pi} \frac{\Im \Pi^{(V)} (s, n)}{s}$.

As usual in QCD sum rules \cite{lit11,lit12}, for large values of $Q^2$ 
one can evaluate 
the r.h.s. of eq.~(\ref{eq_5}) 
by the operator product expansion (OPE) leading to
\bea
\frac{\Pi^{(V)}(Q^2)}{Q^2} = - c_0\; {\rm ln}(Q^2) + \sum\limits_{i=1}^{\infty}
\;\frac{c_i}{Q^{2i}}\;,
\label{eq_15}
\eea
where the coefficients $c_i$ include the Wilson 
coefficients and the expectation values of the corresponding products of the 
quark and gluon field operators, i.e. condensates.

Performing a Borel transformation  
of the dispersion relation (\ref{eq_10}) with appropriate 
parameter $M^2$ and taking into account the 
OPE (\ref{eq_15}) one gets the basic QSR equation 
\bea
\Pi^{(V)} (0,n) + \int\limits_0^{\infty} d s \,R^{(V)} (s)\, {\rm e}^{-s/M^2} = 
c_0 M^2 + \sum\limits_{i=1}^{\infty} \frac{c_i}{(i-1)! M^{2 (i-1)}}\,.
\label{eq_20}
\eea
The advantage of the Borel transformation is 
(i) the exponential suppression of the high-energy part of $R^V (s)$, and 
(ii) the possibility to suppress higher-order
contributions to the r.h.s.\ sum. Choosing sufficiently large values
of the internal technical parameter $M$ one can truncate the sum
in controlled way, in practice at $i = 3$. 
The general structure of the coefficients $c_i$ up to $i=3$ is given, for 
instance, in \cite{lit11,lit19,50_percent}. 

\subsection{QCD condensates} 

In order to calculate the density dependence of the 
condensates entering the coefficients $c_i$ we employ the standard 
linear density approximation, which is valid for not too large density. 
This gives for the chiral quark condensate 
$\langle \overline{q} q\rangle_n = \langle \overline{q} q\rangle_0 
+ \frac{\sigma_N}{2 m_q} n\;$,
where we assume here isospin symmetry for the light quarks, 
i.e. $m_q=m_u=m_d=5.5$ MeV and 
$\langle \bar q q \rangle_0 = \langle \bar u u \rangle_0 = 
\langle \bar d d \rangle_0 = - (0.24 \, {\rm GeV})^3$.
The nucleon sigma term is $\sigma_N = 45 $ MeV.
The gluon condensate is obtained as usual employing the QCD trace anomaly
$\langle\frac{\alpha_s}{\pi} {\rm G^2}\rangle_n = 
\langle\frac{\alpha_s}{\pi} {\rm G^2}\rangle_0 - \frac{8}{9} M_N^0 \;n \;,$
where $\alpha_s=0.38$ is the QCD coupling constant and $M_N^0=770 $ MeV 
is the nucleon mass in the 
chiral limit. The vacuum gluon condensate is 
$\langle \frac{\alpha_s}{\pi} G^2\rangle_0 = (0.33\,{\rm GeV})^4$.

The coefficient $c_3$ in eq.~(\ref{eq_20}) contains also the  
mass dimension-6 4-quark condensates
(cf.\ \cite{Faessler} for a recent calculation of corresponding
matrix elements)  
$\langle (\bar q \gamma_{\mu}\lambda^{a} q)^2 \rangle_n$,
$\langle (\bar u \gamma_{\mu}\lambda^{a} u)(\bar d \gamma^{\mu}\lambda^{a} d)\rangle_n$,
$\langle (\bar q \gamma_{\mu}\lambda^{a} q)(\bar s \gamma^{\mu}\lambda^{a} s)\rangle_n$,
and 
$\langle(\overline{q}\gamma_{\mu}\gamma^5 \lambda^{a} q)^2\rangle_n$
which are common for $\rho$ and $\omega$ mesons.
On this level, $\rho$ and $\omega$ mesons differ only by the condensate
$\pm 2 \langle (\bar u \gamma_\mu \gamma_5 \lambda^a u)
(\bar d \gamma^\mu \gamma_5 \lambda^a d)\rangle_n$ (cf.\ \cite{50_percent}),
causing the small $\rho - \omega$ mass splitting in vacuum \cite{SVZ_2}.
The standard approach to estimate the density dependence of the 4-quark condensates
consists in the use of the mean-field approximation. 
Within such an approximation the 4-quark condensates 
are proportional to $\langle\overline{q} q\rangle_n^2$ 
and their density dependence is actually 
governed by the square of the chiral quark condensate. 
Keeping in mind the important role of the 4-quark condensate for the in-medium 
modifications of the vector mesons, we go beyond this  
approximation and employ the following parameterization \cite{our1}
\be
\langle(\bar q \gamma_\mu \gamma^5 \lambda^a q)^2 \rangle_n =
\frac{16}{9} \langle\overline{q} q\rangle_0^2 \; \hat \kappa_0 \;
\left[1+\frac{\hat \kappa_N}{\hat \kappa_0}\frac{\sigma_N}{m_q 
\langle \bar q q \rangle_0}\;n\right]\;.
\label{eq_35}
\ee 
In vacuum, $n=0$, the parameter $\hat \kappa_0$ 
reflects a deviation from the vacuum saturation assumption. The case
$\hat \kappa_0=1$ corresponds obviously to the exact vacuum saturation \cite{Cohen} 
as used, for instance, 
in \cite{lit19}. To control the deviation of the in-medium 4-quark condensate 
from the mean-field approximation we introduce the parameter $\hat \kappa_N$. 
The limit $\hat \kappa_N = \hat \kappa_0$ recovers the mean-field approximation, while the 
case $\hat \kappa_N > \hat \kappa_0$ ($\hat \kappa_N < \hat \kappa_0$) is 
related to the stronger (weaker) 
density dependence compared to the mean-field approximation. 

An analog procedure applies for the other 4-quark condensates,
each with its own $\hat \kappa_0$ and $\hat \kappa_N$,
which sum up to a parameter $\kappa_0$ and a parameter $\kappa_N$.
Below we vary the poorly constrained parameter $\kappa_N$ to estimate 
the contribution of the 4-quark 
condensates to the QSR with respect to the main trends of the in-medium 
modification of the vector meson spectral function.
As seen in eq.~(\ref{eq_35}) and eq.~(\ref{eq_40}) below,
$\kappa_N$ parameterizes the density dependence of the summed 4-quark condensates;
$\kappa_0$ is adjusted to the vacuum masses. Strictly speaking, $\kappa_0$ and
$\kappa_N$ differ for $\rho$ and $\omega$ mesons due to contributions 
of the above mentioned flavor-mixing condensate; in addition, in medium a twist-4
condensate make further $\rho$ and $\omega$ to differ 
\cite{lit11,50_percent}.
However, the differences can be estimated to be sub-dominant. Therefore, we use
in the present work one parameter $\kappa_N$, keeping in mind that it may slightly
differ for different light vector mesons. 

Using the above condensates and usual Wilson coefficients one gets as relevant terms
for mass dimension $\le 6$ and twist $\le 2$  
\bea
c_0 &=& \frac{1}{8 \pi^2} \left(1 + \frac{\alpha_s}{\pi}\right)\;, \label{eq_2.6}\\
c_1 &=& - \frac{3 m_q^2}{4 \pi^2} \;,\\
c_2 &=& m_q \langle\overline{q} q\rangle_0 + \frac{\sigma_N}{2} \;n + 
\frac{1}{24} \left[\langle\frac{\alpha_s}{\pi} G^2 \rangle_0 
- \frac{8}{9} M_N^0 \;n\right]
+ \frac{1}{4} A_2 M_N \;n\;, \\
c_3 &=& - \frac{112}{81} \pi \;\alpha_s \;\kappa_0\; 
\langle\overline{q} q\rangle_0^{2} 
\left[1+\frac{\kappa_N}{\kappa_0}\frac{\sigma_N}{m_q 
\langle \overline{q} q\rangle_0}\;n\right]  
- \frac{5}{12} A_4 M_N^3 \;n.
\label{eq_40}
\eea
The last terms in $c_{2,3}$ correspond to the derivative condensates 
from non-scalar operators as a consequence of the breaking of Lorentz 
invariance in the medium. These condensates are proportional to the 
moments $A_i=2\int\limits_0^1 d x \;x^{i-1} \left[q_N (x , \mu^2 + 
\overline{q}_N (x, \mu^2))\right]$ of quark and anti-quark distributions inside 
the nucleon at scale $\mu^2=1 {\rm GeV}^2$ 
(see for details \cite{lit11}). Our choice of the 
moments $A_2$ and $A_4$ is $1.02$ and $0.12$, respectively.

The value of $\kappa_0$ in eq.~(\ref{eq_35}) is related 
to such a choice of the chiral condensate $\langle\overline{q} q\rangle_0$ to 
adjust the vacuum vector meson masses. 
In our QSR we have used $\kappa_0=3$, obtaining 
$m_{\rho, \omega}(n=0)=777$ MeV close to the nominal
vacuum values. 
The ratio $\kappa_N/\kappa_0$ in the parameterization (\ref{eq_35}) is restricted by 
the condition 
$\langle(\overline{q} \gamma_{\mu} \lambda^a q)^2\rangle_n \le 0$,   
so that one gets 
$0\le \kappa_N \le 4$ as reasonable numerical limits when considering
$n \le n_0$, as dictated by our low-density approximation.

The case of finite baryon density \underline{and} temperature has been
considered in \cite{our1}.
Here we focus on density effects with the reasoning that temperature
effects below 100 MeV are negligible. 

\subsection{Vector meson spectral density} 

To model the hadronic side of the QSR (\ref{eq_20}) we make the standard 
separation of the vector meson spectral density $R^{(V)}$ into resonance part 
and continuum contribution by means of the threshold parameter $s_V$
\bea
R^{(V)}(s, n)= F_V \;\frac{S^{(V)} (s,n)}{s} \;\Theta(s_V-s) + 
c_0\; \Theta (s-s_V)\;,
\label{eq_45}
\eea
where $S^{(V)} (s,n)$ stands for the resonance peak in the spectral function;
the normalization $F_V$ is 
unimportant for the following consideration. 
In the majority of the previous QCD sum rule evaluations, the zero-width approximation 
\cite{lit11} or some parameterization of $S^{(V)}$ 
\cite{lit13} are employed.
In contrast to this, we use here a more 
realistic ansatz for the resonance spectral density $S^{(V)}$ based on the 
general structure of the in-medium vector meson propagator 
\bea
S^{(V)} (s, n) = - \frac{ \Im \, \Sigma_V (s,n)}
{(s-\stackrel{\rm o}{m}_V^2 (n) - \Re\, 
\Sigma_V (s,n))^2 + 
(\Im \,\Sigma_V(s,n))^2}\;,
\label{eq_50}
\eea
with $\Re \,\Sigma_V(s,n)$ and $\Im \,\Sigma_V(s,n)$ 
as real and imaginary part of the in-medium vector meson 
self-energy. An important point of our approach is that the meson 
mass parameter $\stackrel{\rm o}{m}_V (n) $ becomes 
density dependent in nuclear matter. This dependence is determined by 
the QCD sum rule eq.~(\ref{eq_20}) and mainly governed by the QCD condensates. 
As a result (see below) the in-medium change of the QCD condensates causes  
global modifications of the vector meson spectral function, in 
addition to the collision broadening. (An analogous approach was 
used in \cite{lit20}.) 
The in-medium vector meson mass is determined by the pole position of the 
meson propagator 
\bea
m^2_V (n) = \stackrel{\rm o}{m}_V^2 (n) + \Re \, 
\Sigma_V (s=m_V^2 (n), n)\;,
\label{eq_55}
\eea
which looks similar to the vacuum case, where $n = 0$.
The difference $\Delta m_V(n) \equiv m_V(n)-m_V(0)$ can 
be associated 
with the in-medium vector meson mass shift that is widely used to characterize the 
spectral change of mesons in matter.

Within the linear density approximation the vector meson self energy is 
given by 
\bea
\Sigma_V (E,n) = \Sigma^{\rm vac}_V (E) - n\;T_{V N} (E)\;,
\label{eq_60}
\eea
where $E=\sqrt{s}$ is the meson energy, $\Sigma_V^{\rm vac} (E) 
= \Sigma_V (E,n=0)$, and $T_{V N} (E)$ is the (off-shell) forward 
meson-nucleon scattering amplitude in free space. 
The renormalized quantity  $\Sigma_\rho^{\rm vac}$ 
is summarized in the Appendix A;
for the $\omega$ meson we absorb as usual
$\Re \Sigma_\omega^{\rm vac}$ in $\stackrel{\rm o}{m_\omega}^2$
(cf.\ \cite{Klingl_Kaiser_Weise} for details)
and put $\Im \Sigma_\omega^{\rm vac} = - m_\omega \Gamma_\omega \Theta (E - 3 m_\pi)$
with the vacuum values of mass $m_\omega$ and width $\Gamma_\omega$.

The described framework is well defined, supposed $T_{VN}$ is reliably known.
Unfortunately, the determination of $T_{VN}$ is hampered by uncertainties
(cf.\ results in \cite{lit14} and \cite{lit17}). $\Im T_{VN}$ is more directly
accessible, while $\Re T_{VN}$ follows by a dispersion relation with
sometimes poorly known subtraction coefficients. Since our emphasis here is
to include the collision broadening and finite width effects in the spectral
function, we absorb in the following $\Re T_{VN}$ in 
$\stackrel{\rm o}{m}_V^2(n)$
thus neglecting a possible strong energy dependence. In such a way, the 
uncertainties of $\Re T_{VN}$ become milder since $m_V (n)$ is then mainly 
determined by the QSR. 

We take the needed $\Im T_{V N} (E)$ for $\rho$ 
and $\omega$ mesons from results of the detailed analysis of pion- 
and photon-nucleon scattering data performed recently in \cite{lit17} 
on the footing of the Bethe-Salpeter equation approach with four-point 
meson-baryon contact interactions and a unitary condition for the coupled 
channels. Because of the presence of dynamically generated nucleon resonances, 
like the s-waves N$(1535)$, N$(1650)$ and d-wave N$(1520)$ resonances,
the vector meson-nucleon 
scattering amplitudes obtained in \cite{lit17} exhibit rapid variations with 
energy (see Appendix B, Figs.~\ref{fig_B1} and \ref{fig_B2}). For the $\rho$N channel, 
the dominant contribution in $\Im T_{\rho {\rm N}} (E)$ comes 
from the resonances N$(1535)$ and N$(1520)$. 
Due to the rather moderate coupling of the $\rho N$ channel to 
N$(1520)$, the value of the inelastic $\rho N$ scattering amplitude 
is comparatively small and, therefore, the $\rho$ meson width is not
significantly increased. At the same 
time, N$(1520)$ is coupled strongly to the $\omega N$ channel. 
This causes the pronounced peak in the subthreshold region of 
$\Im T_{\omega {\rm N}} (E)$.
Such a peak like energy dependence 
differs even qualitatively from results of the chiral Lagrangian approach 
\cite{lit14}. 
We do not advocate here a particular effective Lagrangian approach for the vector 
meson-nucleon scattering amplitudes in vacuum. 
Our aim is rather to demonstrate the impact of the QCD side, 
in particular of the in-medium 4-quark condensate, on the 
global vector meson spectral change in nuclear matter. 

For the subtraction constants $\Pi^{(V)} (0,n)$ in eq.~(\ref{eq_10}) 
we use $\Pi^{(\rho)} (0,n) = n/(4 M_N)$, 
$\Pi^{(\omega)} (0,n) = 9 n/(4 M_N)$, 
which are actually the Thomson limit of the 
$VN$ scattering processes, but also coincide with Landau damping terms elaborated 
in \cite{lit18} for the hadronic spectral function entering the dispersion 
relation without subtractions. 
For details about the connection of subtraction constants 
and Landau damping term we refer the interested reader to \cite{lit21}.

\subsection{QCD sum rule} 

Taking the ratio of the eq.~(\ref{eq_20}) to its derivative with respect to 
$M^2$, and using (\ref{eq_45}) one gets 
\begin{eqnarray}
&& \hspace*{-2cm} \frac{\int\limits_0^{s_V} ds \; S^{(V)} (s,n)\;{\rm e}^{-s/M^2}}
{\int\limits_0^{s_V} ds \; S^{(V)} (s,n) \,s^{-1} \; {\rm e}^{-s/M^2}} = \nonumber \\
&& \hspace*{1cm} 
\frac{\displaystyle c_0\,M^2\,[1-\left(1+\frac{s_V}{M^2}\right) {\rm e}^{-s_V/M^2}] 
- \frac{c_2}{M^2} - \frac{c_3}{M^4}}{\displaystyle c_0\,\left(1-{\rm e}^{-s_V/M^2}\right) 
+ \frac{c_1}{M^2} + \frac{c_2}{M^4} + \frac{c_3}{2 M^6} 
- \frac{\Pi^{(V)} (0,n)}{M^2}}
\label{eq_70}
\end{eqnarray}
with the coefficients $c_1, \cdots, c_3$ from eqs.~(\ref{eq_2.6} $\cdots$ \ref{eq_40}) 
and the resonance spectral function $S^{(V)} (s,n)$ from (\ref{eq_50}).
Eq.~(\ref{eq_70}) determines the mass parameter 
$\stackrel{\rm o}{m}_V (n; M^2, s_V)$ being here the subject of the QCD sum rule.


\section{Results of QSR evaluation for $\rho, \omega$ mesons}

Before coming to the results we have to specify the numerical evaluation of the
QCD sum rule (\ref{eq_70}). 

\subsection{Evaluation of the sum rule} 

At a given baryon density $n$ the continuum threshold $s_V$  
is determined by requiring maximum flatness of 
$\stackrel{\rm o}{m}_V (n; M^2,s_V)$ as a function of $M^2$ within 
the Borel window $M_{\rm min}^2\; \cdots \;M_{\rm max}^2$. The minimum Borel 
parameter $M_{\rm min}^2$ is determined such that the terms of order ${\cal O} (M^{-6})$
on the OPE side eq.~(\ref{eq_20}) contribute not more that 10\% \cite{lit13,lit22}.
Selecting such sufficiently large values of  $M_{\rm min}^2$ suppresses higher-order
contributions in the OPE eq.~(\ref{eq_20}) and justifies the truncation.
Typically, $M_{\rm min}^2(10\%)$ is in the order of 0.6 GeV${}^2$.
The values for $M_{\rm max}^2$ are roughly determined by the ''50\% rule''
\cite{50_percent,lit22}, i.e., 
the continuum part of the hadronic side must not contribute
more than 50\% to the total hadronic side. 
According to our experience \cite{our1}, $\stackrel{\rm o}{m}_V$ is not 
very sensitive to variations of $M_{\rm max}^2$. We can, therefore,  
fix the maximum Borel parameter by $M_{\rm max}^2 = 1.5 \,(2.4) \,{\rm GeV}^2$
for the $\omega$ ($\rho$) meson, in good agreement with the ''50\% rule''. 
The sensitivity of the results on these choices of the Borel window
is discussed in Appendix C. Two examples of $\stackrel{\rm o}{m_V}$
as a function of the Borel parameter $M^2$ are displayed in Figs.~\ref{fig_1} and \ref{fig_2}
for our default parameters and for $n = n_0 = 0.15$ fm${}^{-3}$. One observes, indeed,
flat curves $\stackrel{\rm o}{m_V} (n; M^2, s_V)$ within the Borel window.
This is a prerequisite for the stability of the following analyses.

To get finally the vector meson mass parameter  
$\stackrel{\rm o}{m}_V (n)$ we average 
the quantity $\stackrel{\rm o}{m}_V (n; M^2, s_V)$ 
within the above Borel window to get
$\stackrel{\rm o}{m}_V (n) = (M_{\rm max}^2 - M_{\rm min}^2)^{-1}
\int_{M_{\rm min}^2}^{M_{\rm max}^2} dM^2  \, \stackrel{\rm o}{m}_V (n; M^2, s_V)$
which is used in eqs.~(\ref{eq_50}, \ref{eq_55}).

\subsection{In-medium modifications of $\rho, \omega$ masses} 

The results of our QSR evaluations for the density dependence of the vector 
meson masses $m_{\rho} (n)$ and $m_{\omega} (n)$, defined in eq.~(\ref{eq_55}),
for $\kappa_N=1 \cdots 4$ 
are exhibited in Figs.~\ref{fig_3} and \ref{fig_4}, respectively. 
As seen in Fig.~\ref{fig_3} the 
$\rho$ meson mass drops with increasing nucleon density. The value of 
the $\rho$ meson mass shift at given density is directly governed by the 
parameter $\kappa_N$, i.e. the strength of the density dependence of the 
4-quark condensate. Some qualitative arguments to understand 
such an important role of the 4-quark condensate for the 
in-medium $\rho$ meson mass shift are given in \cite{our1}.


The impact of the 4-quark condensate is more pronounced for the isoscalar 
channel. In Fig.~\ref{fig_4} one can observe that such a global characteristic 
as the sign of the $\omega$ meson mass shift is changed by a variation of the 
parameter $\kappa_N$. Similar to the $\rho$ meson the density dependence 
of the $\omega$ meson mass $m_{\omega} (n)$ is mainly governed by the 
QCD mass-parameter $\stackrel{\rm o}{m}_{\omega} (n)$ in accordance with the 
in-medium change of the 4-quark condensate. 
(This confirms previous results obtained within the 
zero-width approximation, which is equivalent to an evaluation of a normalized
moment of the spectral function \cite{our1}, and the finite width treatment in 
\cite{our2} based on an effective chiral Lagrangian \cite{lit14}.) 
In particular, for a weak dependence of the 4-quark condensate on density 
$(\kappa_N \stackrel{<}{\sim} 2)$ the $\omega$ meson mass $m_{\omega} (n)$
is increased, while for a greater value of $\kappa_N$ the $\omega$ meson 
mass decreases with density. 

The sign of the $\omega$ meson mass shift is important with respect to the 
expectation to produce nuclear bound states of $\omega$ meson using suitable 
projectiles impinging on a nuclear target \cite{lit14}. From our study one 
can conclude that a negative $\omega$ meson mass shift, corresponding an effective 
attractive potential, is caused by a  strong dependence of the 4-quark 
condensate on density, i.e. for $\kappa_N \stackrel{>}{\sim} 3$.

The different behavior of $m_\rho (n)$ and $m_\omega (n)$ can be traced back,
to some extent, to different values of the subtraction constants
$\Pi^{(\rho, \omega)} (0, n)$, as emphasized in \cite{lit18}.
The strikingly different vacuum widths, $\Im \Sigma^{\rm vac}_{\rho, \omega}$,
cause further differences, in medium additionally amplified by different
shapes of $T_{\rho, \omega N}$.

Our calculations also show that the main pattern 
of the behavior of $m_V (n)$ plotted in Figs.~\ref{fig_3} and \ref{fig_4} 
remains stable even for the extreme cases when including
$\Re T_{\omega N}$ or discarding $T_{\omega N}$ at all.
This still points to the crucial role of the 4-quark condensate 
for the $\rho, \omega$ meson in-medium mass shifts.
The robustness of the pattern of $m_V (n)$ as a function of the density
under variations of $T_{VN}$ can be interpreted as stringent impact of the
density dependence of the condensates, while the influence of the strong
interaction encoded in $T_{VN}$ is, within the QCD sum rule approach, of
sub-leading order, for the given examples.

\subsection{Spectral functions} 

While the global mass shifts of the in-medium $\rho$ and $\omega$ mesons are 
governed mainly by the strength of the 4-quark condensate density dependence, 
the details of the vector meson spectral functions depend 
also on the meson-nucleon scattering amplitude $T_{VN}$. In Fig.~\ref{fig_5} we plot the 
$\rho$ meson spectral density for $\kappa_N=1 \cdots 3$ at normal 
nuclear density. (Note that the spectral functions determine the
emission of di-electrons from the vector meson decays \cite{lit9,lit23}.)
The main trend of the down shift of the $\rho$ meson spectral 
function peak position is in accordance with the dropping 
$\rho$ meson mass obtained above by a stronger density dependence, 
parameterized by larger values of $\kappa_N$.
When the peak of $S^{(\rho)} (E)$ is in the interval $E \simeq 0.4 \cdots 0.6$ GeV,
i.e. for $\kappa_N \simeq 1 \cdots 3$, the width of the spectral function decreases 
as the peak moves to the smaller values of $E$. This is not a surprise if one takes 
into account the energy dependence of $\Im T_{\rho N} (E)$ in the same 
(subthreshold) energy interval (see Fig.~\ref{fig_B1}), where $\Im T_{\rho N} (E)$ 
also drops with decreasing energy. From this one can also conclude that
in a wide region of $\kappa_N$ the $\rho$ meson does not 
undergo drastic collision broadening at normal nuclear density,
in contrast to earlier expectations but in line with \cite{lit17}.


In Fig.~\ref{fig_6} we display the change of the $\omega$ meson spectral function in 
nuclear matter at normal nuclear density for the same parameters $\kappa_N$ 
as for the $\rho$ meson (see Fig.~\ref{fig_5}). The in-medium spectral 
change is still seen to be dominated by the density dependence of the 4-quark 
condensate. The dependence of the peak position on the parameter $\kappa_N$ 
is similar to $m_{\omega} (n)$, namely, for a weak density dependence of 
the 4-quark condensate $(\kappa_N \stackrel{<}{\sim} 3)$ the peak is 
up-shifted compared to vacuum, while for $\kappa_N \stackrel{>}{\sim} 3$
the peak moves to a smaller value of the energy. For $S^{(\omega)} (E)$ 
with up-shifted peak positions the width remains almost constant. This is 
in agreement with the approximately constant value of $\Im T_{\omega N} (E)$ 
in the region $E \stackrel{>}{\sim} 0.8$ GeV (see Fig.~\ref{fig_B2}). When the peak of 
$S^{(\omega)} (E)$ moves to a smaller energy 
(for $\kappa_N \stackrel{>}{\sim} 3$) the width of the $\omega$ meson increases 
moderately, which is caused by the increase of $\Im T_{\omega N} (E)$ 
(see Fig.~\ref{fig_B2}) in the corresponding interval of energy.

The pure hadronic calculation in \cite{lit17} predicts a slight up-shift of the
original $\omega$ peak. This case is reproduced in our approach by 
$\kappa_N \approx 2.7$. However, such a value of $\kappa_N$ delivers a strong
down-shift of the original $\rho$ peak (see Fig.~\ref{fig_5}), at variance to the results
in \cite{lit17}. Otherwise, in contrast to \cite{lit14}, but in agreement with
\cite{lit17}, the $\rho$ width is less affected by in-medium effects; rather
for a strongly decreasing $\rho$ mass the width may even become smaller,
as discussed above.

\section{$\phi$ meson}

The treatment of the $\phi$ meson proceeds along the same strategy 
as presented above. 
The corresponding current operator in eq.~(\ref{eq_5}) is 
$J_\mu = \bar s \gamma_\mu s$ which renders the coefficients $c_{1,2,3}$ to be used
in eq.~(\ref{eq_70} ) into
\bea
c_1 &=& - \frac{3 m_s^2}{4 \pi^2} \;,\\
c_2 &=& m_s \langle \overline{s} s\rangle_0 + y \frac{m_s \sigma_N}{2 m_q} \; n + 
\frac{1}{24} \left[\langle \frac{\alpha_s}{\pi} G^2 \rangle_0 
- \frac89 M_N^0 \;n \right] 
+ \frac12 A_2^s M_N \;n\;, \\
c_3 &=& - \frac{112}{81} \pi \;\alpha_s \;\kappa_0\; 
\langle\overline{s} s \rangle_0^{2} 
\left[1 + \frac{\kappa_N}{\kappa_0}\frac{\sigma_N y}{m_q 
\langle \overline{s} s \rangle_0}\;n\right]  
- \frac56 A_4^s M_N^3 \;n\;,
\label{eq_444}
\eea
where $c_0$ is not changed. At the scale $\mu^2 = 1$ GeV${}^2$ the condensates
are $A_2^s = 0.05$ and $A_4^s = 0.002$ \cite{lit11}.
$y = \langle N \vert \bar s s \vert N \rangle / \langle N \vert \bar q q \vert N \rangle $ 
is the poorly known strangeness content of the nucleon which may
vary from 0 to 0.25 \cite{strange_nucleon}. We utilize here $y =
0.22$, as in \cite{lit11}.
Further parameters are 
$\langle \bar s s \rangle_0 = \hat y \langle \bar q q \rangle_0$
with $\hat y = 0.8$ and
$m_s = 130$ MeV. 
The subtraction constant is negligible, i.e. $\Pi^{(\phi)} (0, n) = 0$ \cite{Asakawa_Ko}.
$\Re \Sigma_\phi^{\rm vac} - n \Re T_{\phi N}$ is absorbed again in 
$\stackrel{\rm o}{m_\phi} (n)$, while 
$\Im \Sigma_\phi^{\rm vac} (E) = - m_\phi \Gamma_\phi \Theta (E - 2 m_K))$
with vacuum parameters $m_K, m_\phi, \Gamma_\phi$. 
$M_{\rm max}^2 = 3$ GeV${}^2$ is dictated by the ''50\% rule''. 

For $\Im T_{\phi N}$ we employ the previous estimates
\cite{lit14} (see solid curve in figure 8 in first reference of \cite{lit14}).
Since this $\Im T_{\phi N}$ is comparatively large we find some weak dependence
of $m_\phi$ on $\kappa_N$, see Fig.~\ref{fig_7}. The pattern of the $\kappa_N$
dependence resembles the one of the $\rho$ meson but is much more moderate.
(Note that the slope of the curves $m_\phi (n)$ scale with $y$ \cite{our3}.)
Since the used amplitude $T_{\phi N}$ shows minor variations at $E \sim m_\phi$,
the widths of the shifted spectral functions is quite independent of $\kappa_N$,
see Fig.~\ref{fig_8}.    
(When using the amplitude of \cite{Klingl} the width would become larger
with increasing values of $\kappa_N$.)

\section{Summary and discussion}

In summary we present a systematic evaluation of the Borel
QCD sum rule for $\rho$ and $\omega$ mesons. We go beyond the often
employed zero-width approximation and use a realistic ansatz for the
spectral function. A crucial element for our analysis is the use of
the recent $\rho, \omega$ meson-nucleon scattering
amplitudes adjusted to a large data basis. These differ noticeably
from earlier employed amplitudes. Despite of such differences, the
results of our analysis are robust: The $\rho$ meson suffers a down shift
by an amount determined by the yet poorly known density dependence of the
4-quark condensates. The latter ones determine also whether the $\omega$
meson suffers an up-shift or a down-shift. One consequence of the  
scattering amplitudes \cite{lit17} is a moderate in-medium broadening of
the $\rho, \omega$ spectral functions, in contrast to earlier
predictions. We focus on the region in the vicinity of the $\rho, \omega$
peaks in vacuum. Therefore, we do not address such problems as the development
of second, low-energy peak in the $\omega$ strength, as found in \cite{lit17}. 

Besides the exploration of the importance of the density dependence of the
4-quark condensates, the determination of the in-medium modification of
$\rho$ and $\omega$ on a common footing is the main objective of the present
paper. This is highlighted in Fig.~\ref{fig_9},
which points to drastic shifts of either the $\omega$ meson or the $\rho$ meson,
or to still noticeable shifts of both. Fairly independent of $\kappa_N$
is the $\rho - \omega$ mass splitting of about 200 MeV
at normal nuclear matter density. 
(Using $\Im T_V$ from \cite{lit14} results in a smaller
$\rho - \omega$ mass splitting which even disappears for
$\kappa_N < 1$.) It should be stressed, however, that the use of a common
parameter $\kappa_N$ for the light vector mesons is an approximation,
since actually $\rho$, $\omega$ and $\phi$ mesons have their own
$\kappa_N$'s. The detailed analysis deserves a separate investigation. 
  
It turns out that the in-medium cross properties of the $\rho$ and $\omega$ mesons
are determined, to a large extent, by the condensates, while the meson-nucleon
scattering amplitudes are important for the quantitative behavior.
(E.g.\ $\Im T_{\rho N}^{[16]} > 
\Im T_{\rho N}^{[19]}$
causes more support of $S^{(\rho)} (\Im T_{\rho N}^{[16]})$
than $S^{(\rho)} (\Im T_{\rho N}^{[19]})$ at smaller values
of $E$, which is compensated by a stronger down-shift of $m_\rho (n)$
when using $\Im T_{\rho N}^{[19]}$. Otherwise,
$\Im T_{\omega N}^{[16]} < 
\Im T_{\omega N}^{[19]}$ for $E < 770$ MeV and 
$\Im T_{\omega N}^{[16]} >
\Im T_{\omega N}^{[19}]$ for $E > 770$ MeV which explains the 
somewhat larger up-shift of $m_\omega (n)$ when using 
$\Im T_{\omega N}^{[19]}$ and small values of $\kappa_N$. 
Directly evident is that different 
$\Im T_{V N}$ can cause different shapes of the spectral function.) 
Basing on this observation we consider also the $\phi$
meson using estimates of the $\phi$ meson-nucleon scattering amplitude.
In contrast to the $\rho, \omega$ mesons, the in-medium
modification of the $\phi$ meson is determined by the strangeness chiral condensate
and depends essentially on the strangeness content of the nucleon.

In our approach we rely on the linear density approximation.
There are examples in the literature (e.g.\ \cite{Lutz_Friman_Appel})
which show that, e.g., the chiral
condensate begins to deviate from the linear density behavior at normal
nuclear matter density. Resting on this argument one can expect the
quantitative validity of our results up to $n_0$.

We have truncated, according to the common praxis, the OPE at order 3.
Higher-order terms are not yet calculated in a systematic way. 
This issue needs further consideration, as also the case of a finite
spatial momentum of the vector mesons \cite{finite_momenta}.

Concerning an experimental opportunity to observe both $\rho$ and $\omega$
mesons in-medium mass shifts simultaneously in heavy-ion collisions,
our analysis still shows the crucial importance of the in-medium
density dependence of the 4-quark condensate. In particular,
if the 4-quark condensate density dependence is not too strong
(i.e. for $\kappa_N \stackrel{\sim}{<} 2$) there is a chance to observe
the up-shifted peak of the $\omega$ resonance, while the $\rho$ meson
is down-shifted.   
The measurements with HADES, once the $\rho$ and $\omega$ peaks are identified,
will constrain the mentioned important density dependence of the 4-quark
condensates and, consequently, the strength of approaching chiral symmetry
restoration. 

\noindent{\em Acknowledgments:} 
We thank E.G. Drukarev, R. Hofmann, S. Leupold, V.I. Zakharov, and 
G.M. Zinovjev for useful discussions.  
We are especially grateful to M.F.M. Lutz and Gy. Wolf 
for discussions on the results of \cite{lit17} and for supplying the
information of $T_{VN}$.
O.P.P. acknowledges the warm hospitality of the nuclear 
theory group in the Research Center Rossendorf. This work is supported 
by BMBF 06DR121, STCU 15a, CERN-INTAS 2000-349, NATO-2000-PST CLG 977 482.


\renewcommand{\theequation}{\Alph{section}.\arabic{equation}}

\appendix

\section{Self-energy of $\rho$ meson in vacuum} 

The self-energy of $\rho$ meson in vacuum is
$
\Sigma^{\rm vac}_{\rho} (q) = \frac{1}{3} \; g_{\mu \nu}\,
\Sigma^{\mu \nu} (q)
$
where the self-energy tensor $\Sigma^{\mu \nu} (q)$ within an effective
Lagrangian for the $\rho \pi \pi$ interaction is given by \cite{lit23}
\bea
i \; \Sigma^{\mu \nu} (q) = g^2_{\rho \pi \pi} \int \frac{d^4 p}{(2 \pi)^4} \;
\frac{1}{[p^2 - m_{\pi}^2 + i \epsilon]}\;
\left(\frac{(2 p - q)^{\mu} \; (2 p - q)^{\nu}}
{[(p - q)^2 - m_{\pi}^2 + i \epsilon]}
- \; 2 g^{\mu \nu} \right)\;
\label{ap2}
\eea
with the coupling constant $g_{\rho \pi \pi} = 5.79$ and $m_{\pi} = 0.138$ GeV.
Using the renormalization scheme of \cite{lit17,lit24} we get for the meson 
at rest, i.e. $q=(E, {\bf 0})$,
\bea
\Sigma^{\rm vac}_{\rho} (E) & = & \frac{g_{\rho \pi \pi}^2}{48 \pi^2}
\;(4 m_{\pi}^2 - E^2) \left\{ 
\sqrt{1 - \frac{4 m_{\pi}^2}{E^2}}\; \ln \left(
\frac{\sqrt{1 - \frac{4 m_{\pi}^2}{E^2}} - 1}
{\sqrt{1 - \frac{4 m_{\pi}^2}{E^2}} + 1} \right) \right. \nonumber\\
&& 
 - \left. \sqrt{1 - \frac{4 m_{\pi}^2}{m_{\rho}^2 (0)}}\; {\rm ln}
\left(\frac{1 - \sqrt{1 - \frac{4 m_{\pi}^2}{m_{\rho}^2 (0)}}}
{1 + \sqrt{1 - \frac{4 m_{\pi}^2}{m_{\rho}^2 (0)}}} \right) \right\}\;,
\label{ap3}
\eea
where $m_{\rho} (0) = 0.769$ GeV is the vacuum mass of $\rho$ meson.
In this scheme, $m_{\rho} (0) = \stackrel{\rm o}{m}_{\rho} (0)$
follows from eq.~(\ref{eq_55}).

\section{$\rho, \omega$ - nucleon scattering amplitudes} 

For definiteness we plot in Figs.~\ref{fig_B1} and \ref{fig_B2} the spin and isospin 
averaged amplitudes for $\rho$ and $\omega$ mesons, respectively, which are 
employed in our QSR evaluations and not explicitly given in \cite{lit17}.

\section{Technical details} 

Here we would like to report a few technical details of our sum rule evaluation.
Let us first consider the density dependence of the continuum threshold,
see Figs.~\ref{fig_C1} and \ref{fig_C2}. When changing the density, but keeping the rules
for the Borel window as described above, the continuum thresholds $s_V$ change.
The overall pattern resembles the behavior of $\stackrel{\rm o}{m_V}$:
a decreasing (increasing) $\stackrel{\rm o}{m_V}$ implies a decreasing (increasing)
$s_V$.

If one would freeze the continuum thresholds to the vacuum values, i.e.,
$s_V(n) = s_V(0)$, the $\rho - \omega$ mass splitting at normal nuclear matter
density is reduced to about 100 MeV and the dependence on $\kappa_N$ becomes
much weaker.  

\begin{figure}[!h]
\includegraphics[scale=1.0]{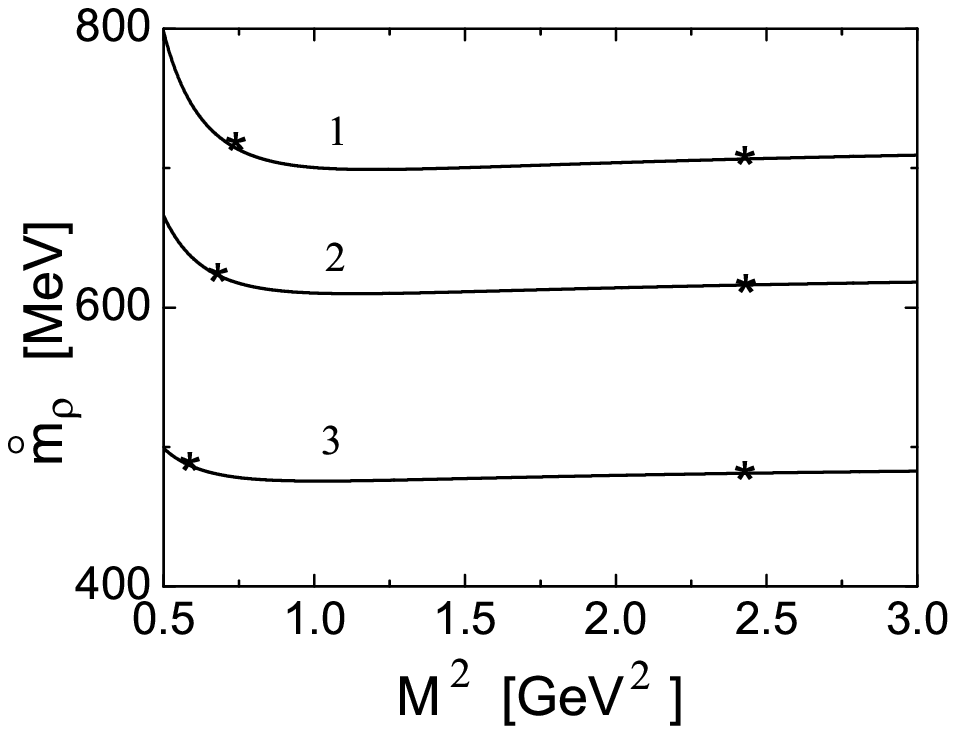}
\caption{
The $\rho$ meson mass parameter $\stackrel{\rm o}{m_\rho}$
as a function of the Borel parameter $M^2$. The respective
continuum thresholds $s_\rho = 1.13, 0.92,$ and 0.65 GeV${}^2$
for  $\kappa_N = 1, 2$ and 3 follow from the maximum flatness requirement
within the Borel window (marked by stars)
defined here by $M^2_{\rm min} (10\%)$ and
$M^2_{\rm max} = 2.4$ GeV${}^2$. $ n = n_0$.}
\label{fig_1}
~\vskip -12mm
\end{figure}

\begin{figure}[!h]
\includegraphics[scale=1.0]{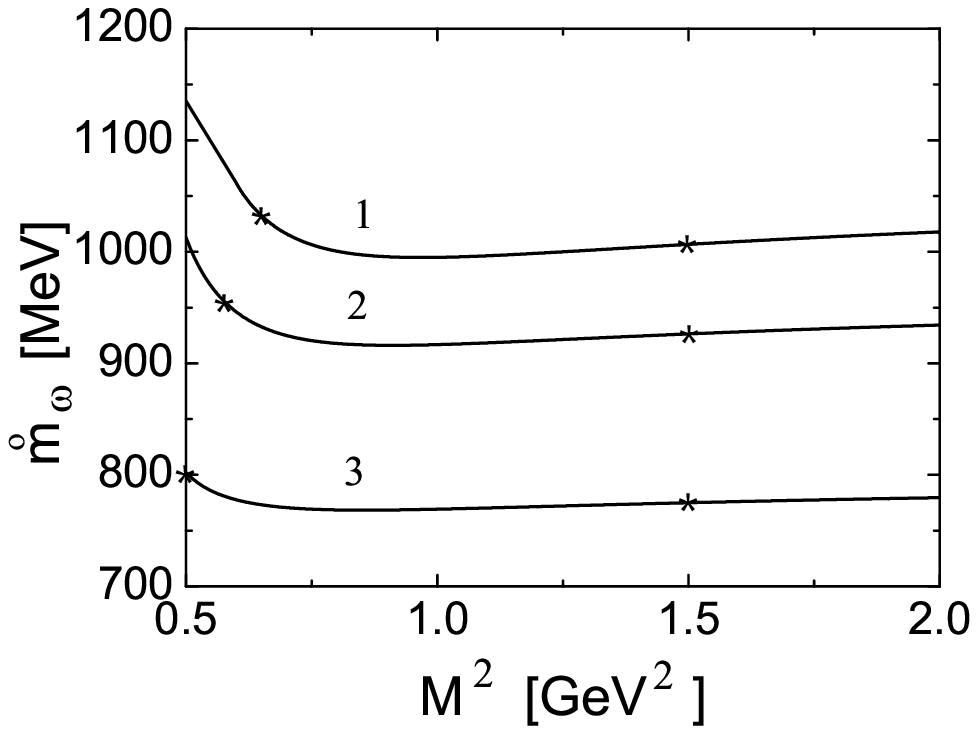}
\caption{
As in Fig.~\protect\ref{fig_1}, but for $\omega$ meson.
$M^2_{\rm max} = 1.5$ GeV${}^2$. }
\label{fig_2}
\end{figure}

\begin{figure}[!h]
\includegraphics[scale=1.0]{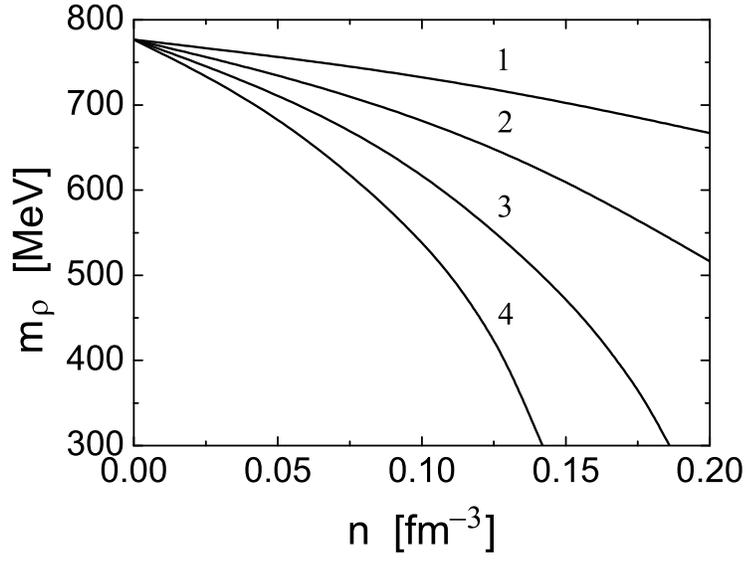}
\caption{Density dependence of the $\rho$ meson mass for 
various values of the parameter $\kappa_N$.}
\label{fig_3}
~\vskip -15mm
\end{figure}

\begin{figure}[!h]
\includegraphics[scale=1.0]{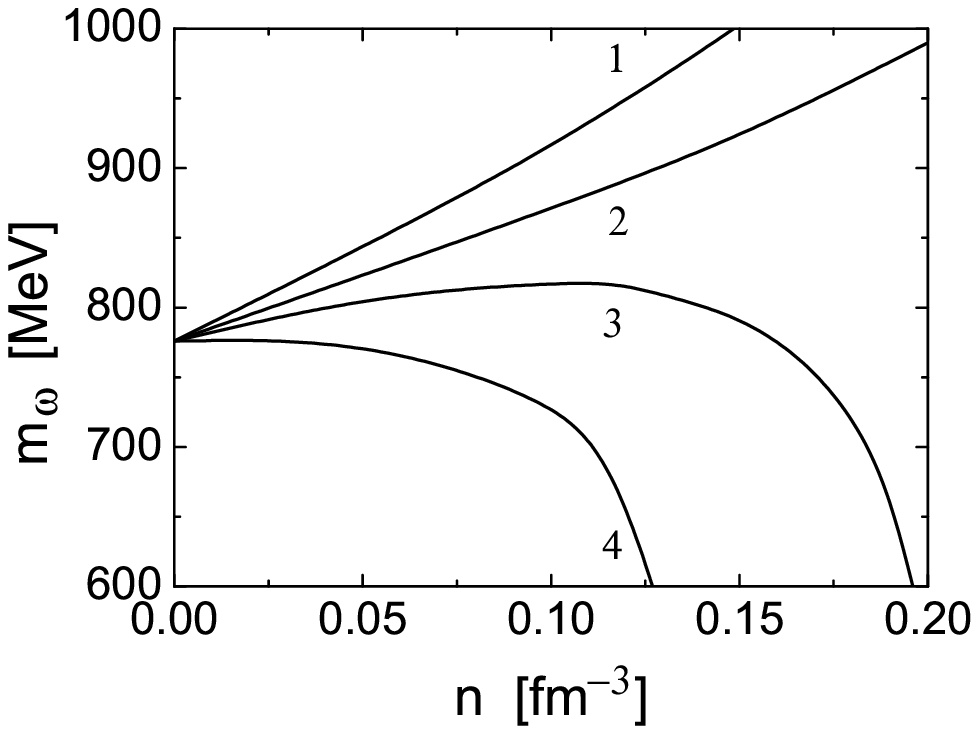}
\caption{As in Fig.~\protect\ref{fig_3}, but for $\omega$ meson.}
\label{fig_4}
\end{figure}

\begin{figure}[!h]
\includegraphics[scale=1.0]{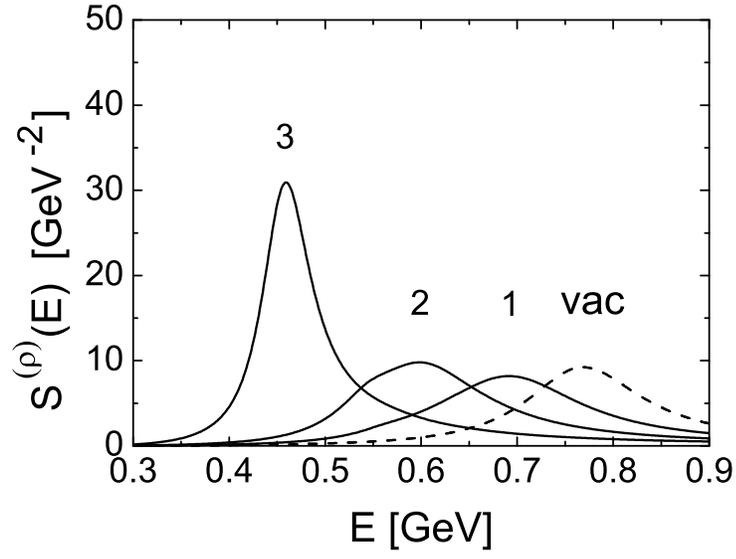}
\caption{$\rho$ meson spectral density for $\kappa_N=1 ... 3$. 
Solid curves correspond to normal nuclear density, 
while the dashed curve is for vacuum.}
\label{fig_5}
\end{figure}

\begin{figure}[!h]
\includegraphics[scale=1.0]{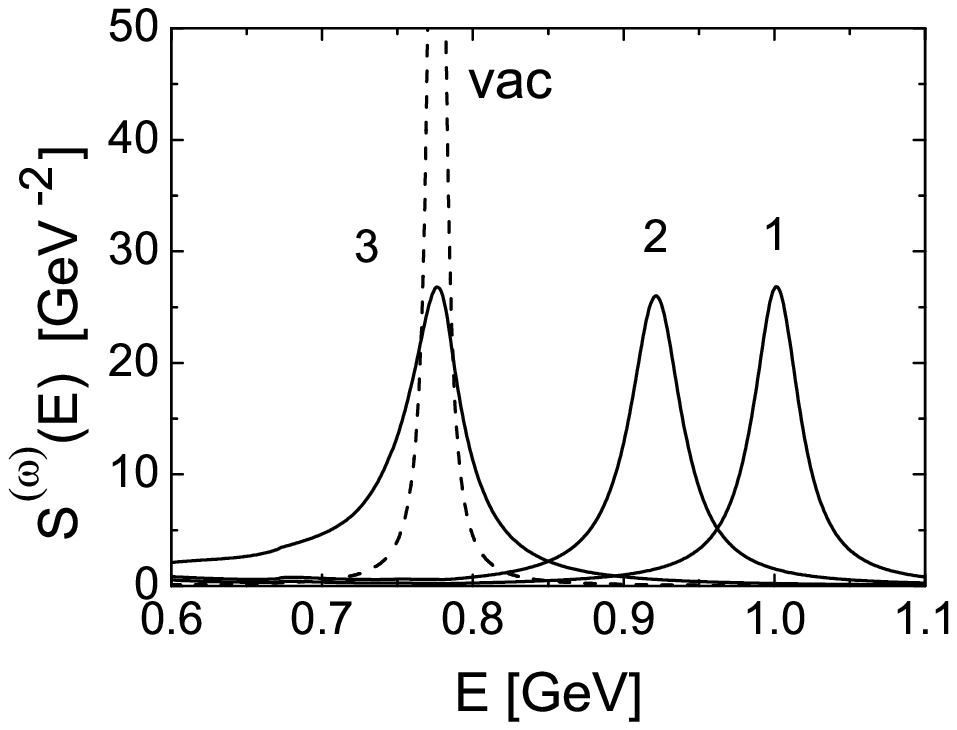}
\caption{As in Fig.~\protect\ref{fig_5}, but for $\omega$ meson.}
\label{fig_6}
\end{figure}

\begin{figure}[!h]
\includegraphics[scale=1.0]{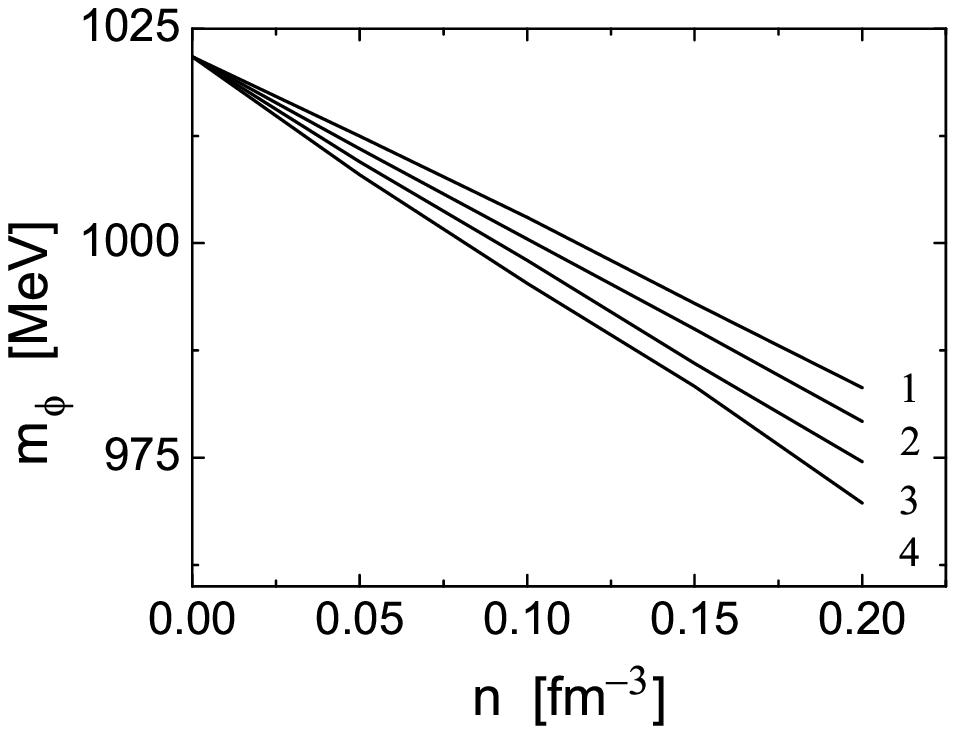}
\caption{
As in Fig.~\protect\ref{fig_3}, but for $\phi$ meson.}
\label{fig_7}
\end{figure}

\begin{figure}[!h]
\includegraphics[scale=1.0]{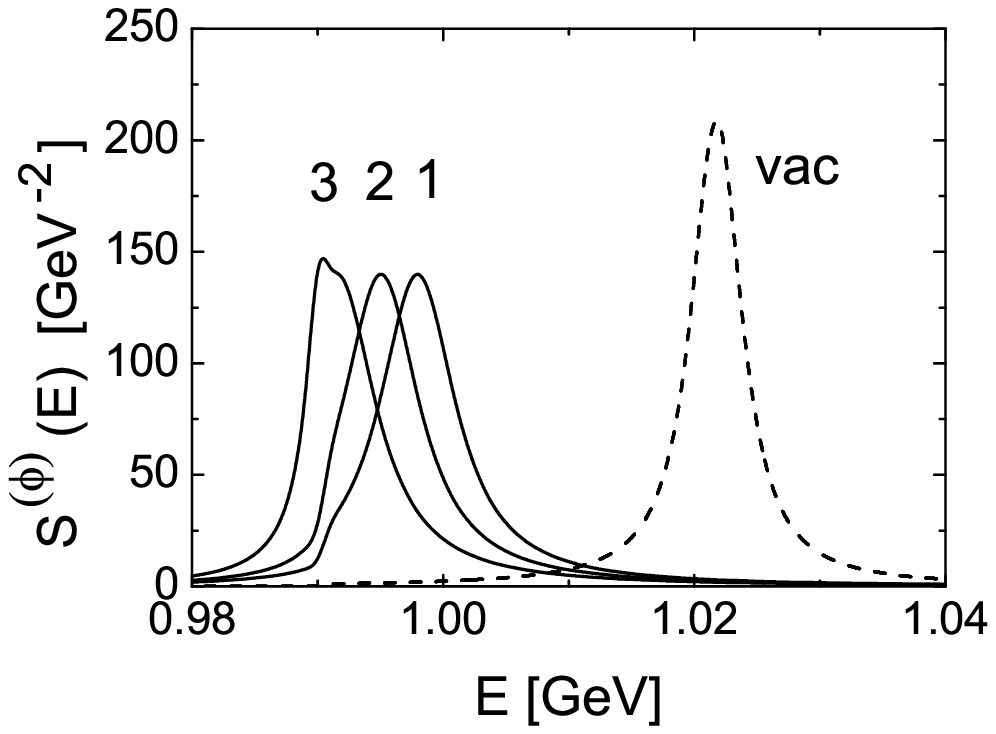}
\caption{
As in Fig.~\protect\ref{fig_5}, but for $\phi$ meson.}
\label{fig_8}
\end{figure}

\begin{figure}[!h]
\includegraphics[scale=1.5]{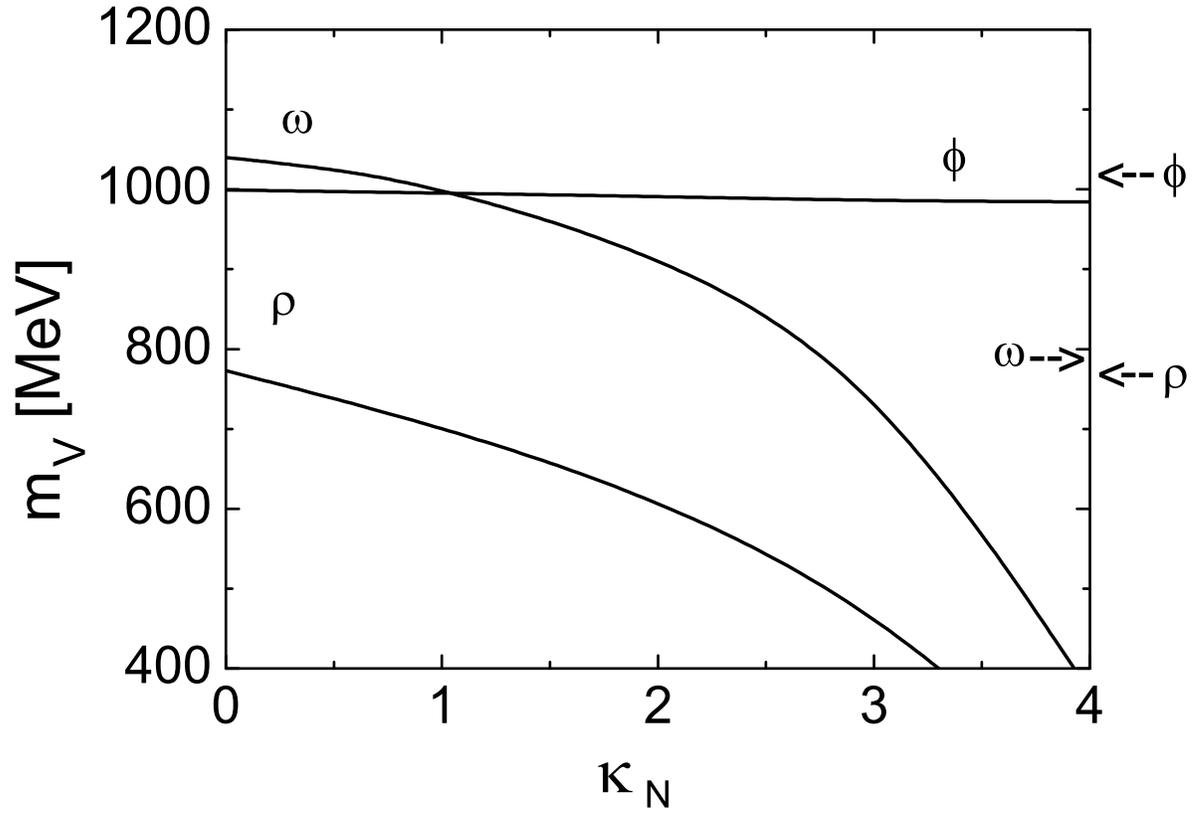}
\caption{
Dependence of the vector meson masses on the parameter $\kappa_N$ at normal nuclear
matter density. The arrows depict the vacuum masses.}
\label{fig_9}
\end{figure}

Next we consider the stability of our results with respect of the choice of the Borel
window at normal nuclear matter density. 
Figs.~\ref{fig_C3} and \ref{fig_C4} exhibit the change of the parameter 
$\stackrel{\rm o}{m_V}$ when changing $M_{\rm min}^2$. 
As expected, $\stackrel{\rm o}{m_V}$ slightly increases
with decreasing $M_{\rm min}^2$ (compare also with Figs.~\ref{fig_1} and \ref{fig_2}). 
The change is fairly
moderate but points to some dependence of the absolute values of 
$\stackrel{\rm o}{m_V}$ and $m_V$ on the Borel window.
This, however, is not important since our
focus here is the pattern of the in-medium modification and not absolute predictions,
which are hampered anyhow by the uncertainty related with the 4-quark condensate.
A similar statement holds for changes of $M_{\rm max}^2$, 
see Figs.~\ref{fig_C5} and \ref{fig_C6}.
With virtue to Figs.~\ref{fig_1} and \ref{fig_2} the decrease of 
$\stackrel{\rm o}{m_V}$ with increasing $M_{\rm max}^2$ is counter-intuitive. 
The explanation of this behavior comes from
the change of $s_V$ when changing $M_{\rm max}^2$. 
When directly determining $M_{\rm max}^2$ by the ''50 \% rule'' \cite{50_percent,lit22}
we arrive at a sliding Borel window where the Borel sum rule
(\ref{eq_20})
is explicitly solved. The results displayed in Fig.~\ref{fig_9} turn
out to be stable.




\newpage


\newpage


\newpage


\newpage


\newpage

\newpage

\begin{figure}[!h]
\includegraphics[scale=1.0]{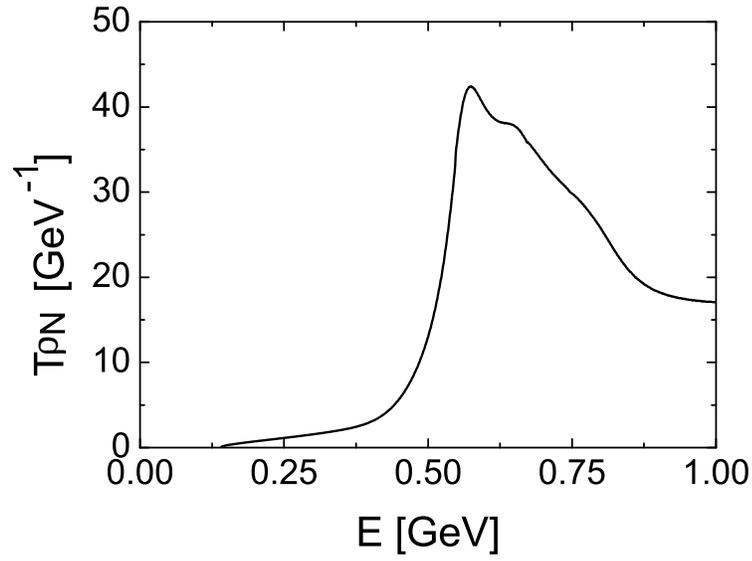}
\caption{
Imaginary part of the
spin and isospin averaged $\rho$ meson-nucleon scattering amplitude
for the approach of \protect\cite{lit17}.}
\label{fig_B1}
\end{figure}
\begin{figure}[!h]
\includegraphics[scale=1.0]{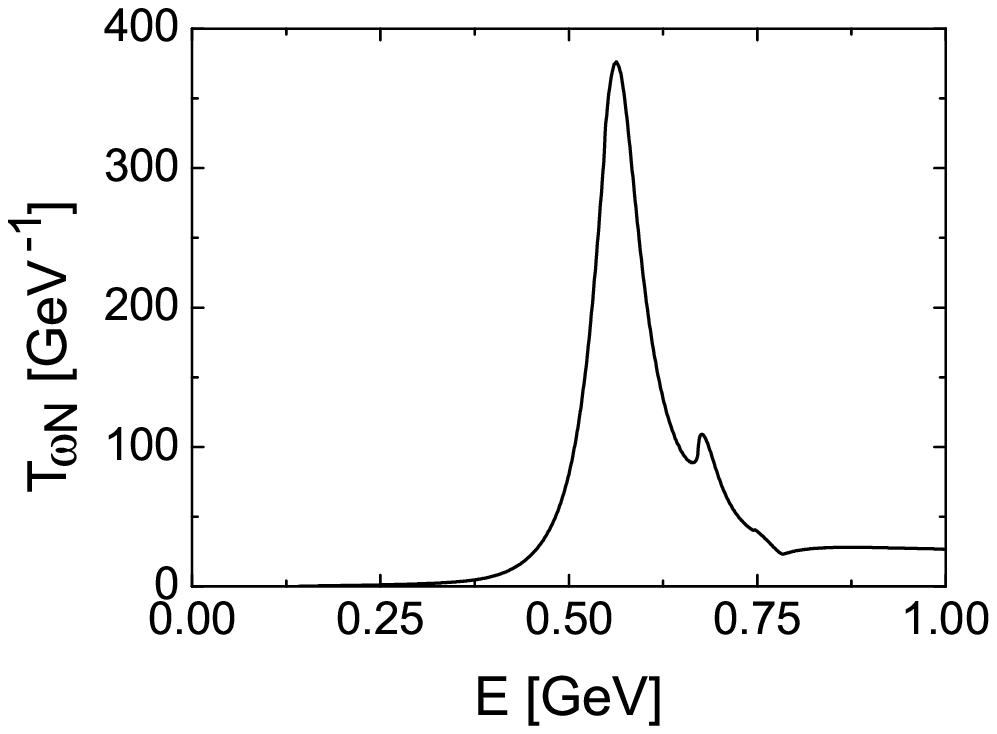}
\caption{
As in Fig.~\protect\ref{fig_B1}, but for $\omega$ meson.} 
\label{fig_B2}
\end{figure}

\newpage

\begin{figure}[!h]
\includegraphics[scale=1.0]{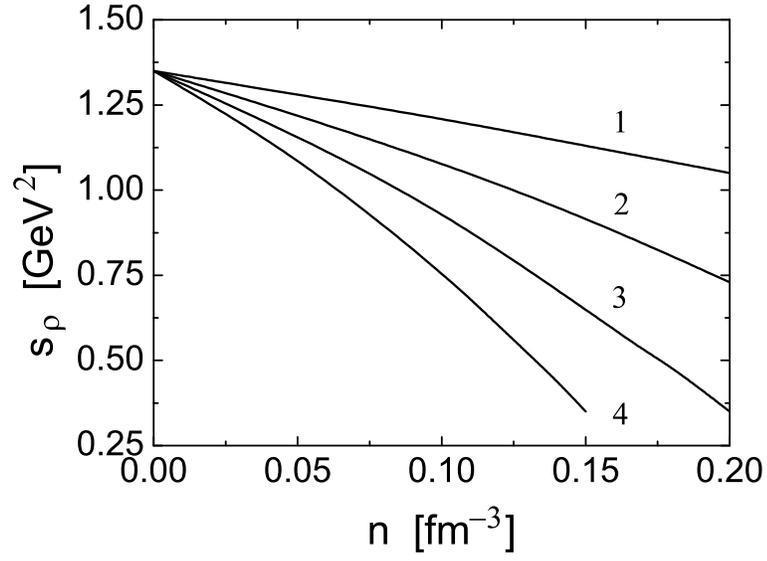}
\caption{ 
Density dependence of the continuum threshold $s_\rho$ for various values
of the parameter $\kappa_N$.} 
\label{fig_C1}
\end{figure}
\begin{figure}[!h]
\includegraphics[scale=1.0]{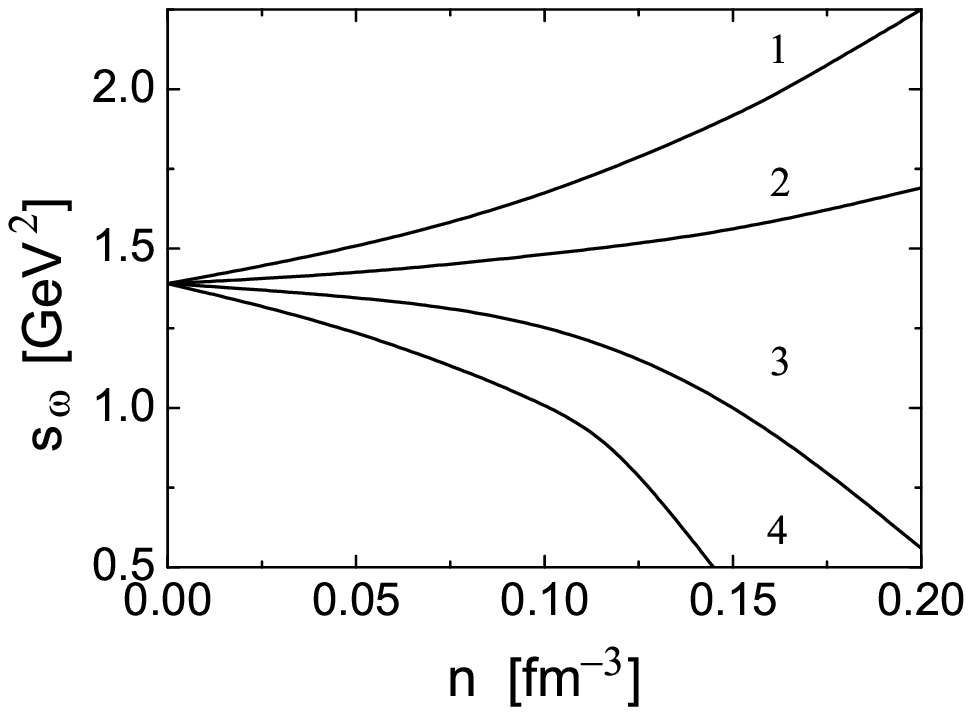}
\caption{
As in Fig.~\protect\ref{fig_C1}, but for $\omega$ meson.} 
\label{fig_C2}
\end{figure}

\newpage

\begin{figure}[!h]
\includegraphics[scale=1.0]{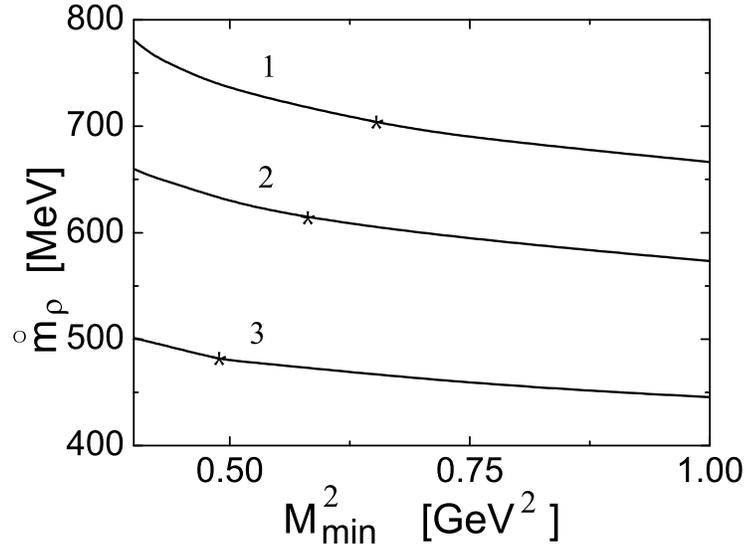}
\caption{
The parameter $\stackrel{\rm o}{m_\rho}$ as a function of the minimum
Borel parameter $M_{\rm min}$. $M^2_{\rm max} = 2.4$ GeV${}^2$.
The stars mark $M_{\rm min}^2 (10\%)$. $n = n_0$.}
\label{fig_C3}
\end{figure}
\begin{figure}[!h]
\includegraphics[scale=1.0]{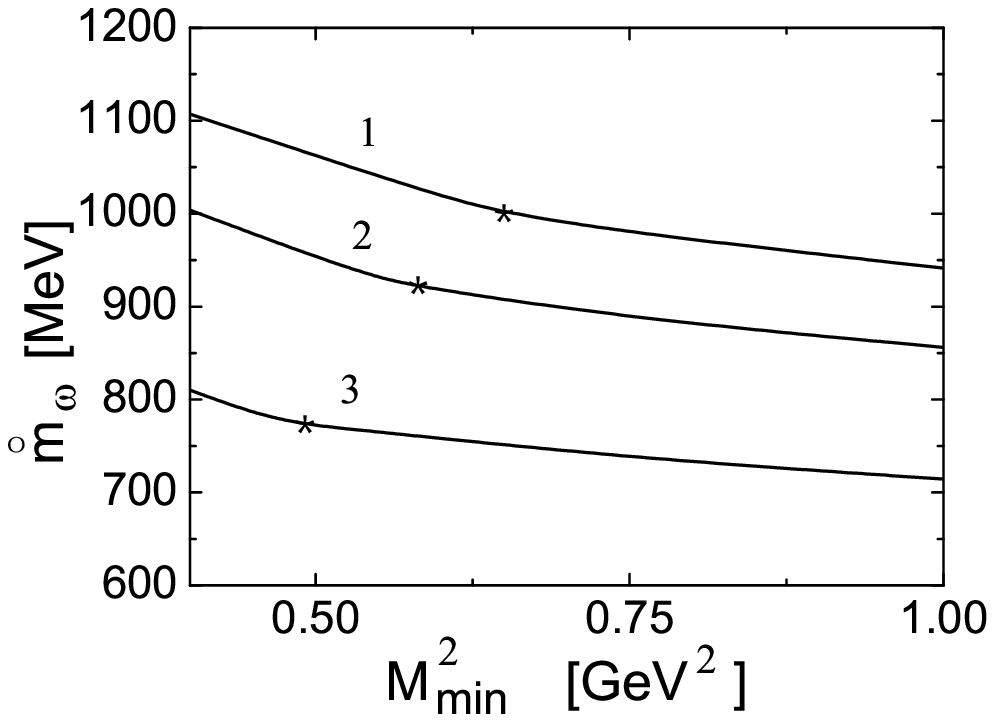}
\caption{
As in Fig.~\protect\ref{fig_C3}, but for $\omega$ meson.
$M^2_{\rm max} = 1.5$ GeV${}^2$.} 
\label{fig_C4}
\end{figure}

\newpage

\begin{figure}[!h]
\includegraphics[scale=1.0]{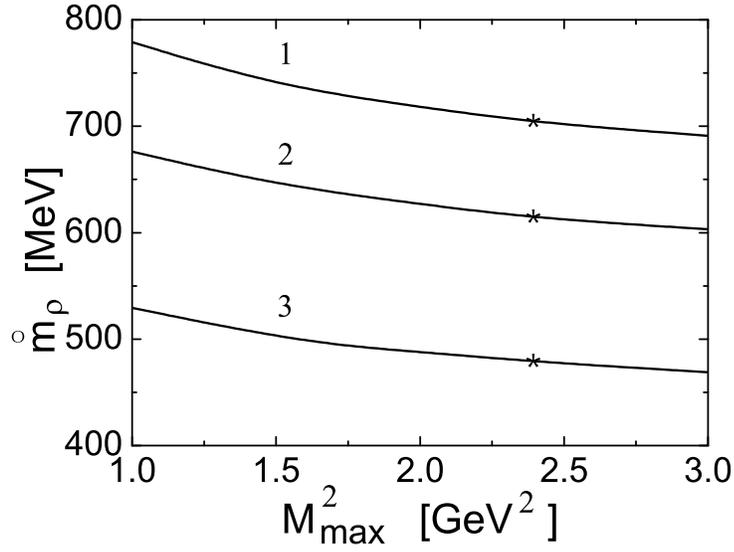}
\caption{
The parameter $\stackrel{\rm o}{m_\rho}$ as a function of the maximum
Borel parameter $M_{\rm max}^2$. 
The minimum Borel parameter is $M_{\rm min}^2 (10\%)$.
The stars mark $M_{\rm max}^2 = 2.4$ GeV${}^2$. $n = n_0$.}
\label{fig_C5}
\end{figure}
\begin{figure}[!h]
\includegraphics[scale=1.0]{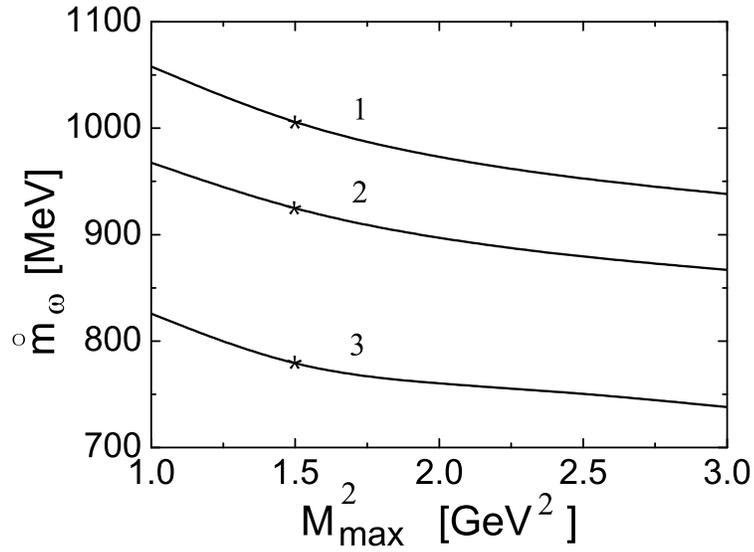}
\caption{
As in Fig.~\protect\ref{fig_C5}, but for $\omega$ meson.
The stars mark $M_{\rm max}^2 = 1.5$ GeV${}^2$.}
\label{fig_C6}
\end{figure}

\end{document}